\documentclass[prb,superscriptaddress,twocolumn,showpacs,floatfix]{revtex4-1}
\usepackage{bm}
\usepackage{amsmath}
\usepackage{amsfonts}
\usepackage{amssymb}
\usepackage{graphicx,graphics,color,
}

\newcommand{\bq}{\begin{equation}}
\newcommand{\eq}{\end{equation}}

\begin{document}

\title{Strong coupling theory of heavy fermion criticality}
\author{Elihu Abrahams}
\affiliation{Department of Physics and Astronomy, University of California
Los Angeles, Los Angeles, CA 90095}
\author{J\"{o}rg Schmalian}
 \affiliation{Institute for Theory of
Condensed Matter, Karlsruhe Institute of Technology, 76049 Karlsruhe,
Germany}
\author{Peter W\"{o}lfle}
 \affiliation{Institute for Theory of
Condensed Matter, Karlsruhe Institute of Technology, 76049 Karlsruhe,
Germany}
\affiliation{Institute for Nanotechnology, Karlsruhe Institute of Technology, 76031 Karlsruhe, Germany}
\date{\today{}}

\begin{abstract}
We present a theory of the scaling behavior of the thermodynamic, transport
and dynamical properties of a three-dimensional metal at an
antiferromagnetic (AFM) critical point. We show how the critical spin
fluctuations at the AFM wavevector $q=Q$ induce energy fluctuations at small 
$q$, giving rise to a diverging quasiparticle effective mass over the whole
Fermi surface. The coupling of the fermionic and bosonic degrees of freedom
leads to a self-consistent relation for the effective mass, which has a
strong coupling solution in addition to the well-known weak-coupling
spin-density-wave solution. We use the recently-introduced concept
of critical quasiparticles, employing a scale-dependent effective mass ratio 
$m^{\ast}/m$ and quasiparticle weight factor $Z$. We adopt a
scale-dependent vertex correction that boosts the coupling of fermions and spin
fluctuations. The ensuing spin fluctuation spectrum obeys $\omega /T$%
-scaling. Our results are in good agreement with experimental data on the
heavy fermion compounds YbRh$_{2}$Si$_{2}$ and CeCu$_{6-x}$Au$_{x}$ for
$3D$ and $2D$ spin fluctuations, respectively.
\end{abstract}

\pacs{}
\maketitle

\affiliation{Department of Physics and Astronomy, University of California
Los Angeles, Los Angeles, CA 90095}

\affiliation{Institute for Theory of
Condensed Matter, Karlsruhe Institute of Technology, 76049 Karlsruhe,
Germany}

\affiliation{Institute for Theory of
Condensed Matter, Karlsruhe Institute of Technology, 76049 Karlsruhe,
Germany} 
\affiliation{Institute for Nanotechnology, Karlsruhe Institute of
Technology, 76031 Karlsruhe, Germany}

\preprint{}

\affiliation{Department of Physics and Astronomy, University of California
Los Angeles, Los Angeles, CA 90095} 
\affiliation{Institute for Theory of
Condensed Matter, Karlsruhe Institute of Technology, 76049 Karlsruhe,
Germany} 
\affiliation{Institute for Nanotechnology, Karlsruhe Institute of
Technology, 76021 Karlsruhe, Germany}

\section{Introduction}

Quantum phase transitions in metallic compounds have attracted considerable
interest over the last two decades. These systems exhibit deviations from
the standard Fermi liquid model. This ``non-Fermi liquid"
behavior is a consequence of the interaction of the fermionic (Landau)
quasiparticles with bosonic critical fluctuations. Early theories \cite%
{Hertz,Millis} of quantum critical behavior, formulated in the framework of
a Ginzburg-Landau-Wilson action of the order parameter field $\phi $, found
that the effective dimension of the corresponding $\phi ^{4}$-field theory
is increased to $d_{eff}=d+z$ where $d,z$ are the spatial dimension of the
fluctuations and the dynamical critical exponent, respectively. In many
cases of interest $d_{eff}$ is above the upper critical dimension, so that
the fluctuations are effectively non-interacting and the theory is of the
Gaussian type (for a review see Ref.\ \onlinecite{LRVW}). While this theory is well
founded in the case of non-metallic systems, for metallic systems the
question arises whether the fermionic degrees of freedom may be easily
integrated out. In this paper, we show that it is often not possible to
reduce the description to that of Hertz and Millis. \cite{Hertz,Millis}
Rather, the interplay of fermionic and bosonic degrees of freedom generates
critical behavior of the fermionic quasiparticles, acting back on the
spectrum of bosonic fluctuations. Then, a strong coupling regime with
respect to the fermion-boson coupling may be reached.

Experimentally well-studied candidate systems we shall focus on are the
heavy-fermion compounds CeCu$_{6-x}$Au$_{x}$ (CCA) for which, guided by
experiment, we consider quasi-two-dimensional antiferromagnetic (AFM) spin
fluctuations at the quantum critical point (QCP) $x=0.1$ \cite{HvL1} and YbRh$_{2}$Si$_{2}$
(YRS), for which we 
assume AFM fluctuations of three-dimensional character at the magnetic field-tuned quantum critical point
 $H=H_{c}$ at temperatures $T$ less than 0.3 K, crossing
over to three-dimensional ferromagnetic fluctuations at higher $T$. \cite{bro} 

An issue that has hampered progress in developing a strong-coupling
theory of criticality that involves AFM fluctuations has been that they
transfer a large momentum of the order of the ordering wave vector ${\bf q}={\bf Q}$. As a consequence, the self-energy in one-loop order becomes highly
anisotropic, being critical only at so-called ``hot spots" on the Fermi
surface that are connected by $\bf{Q}$. However, the combined exchange
of two AFM fluctuations, which may be viewed as a spin-exchange energy
fluctuation, \cite{hart} may transfer only a small momentum $q$ and we argue
here that the one-loop order process of such energy fluctuations is dominant
in providing a renormalization of the fermionic quasiparticle effective mass
that is approximately uniform over the Fermi surface. Thus, we consider here
the simplest case, in which the fermion self energy is only weakly momentum
dependent. A different way by which the effect of the critical AFM
fluctuations may be distributed all over the Fermi surface is by means of
impurity scattering. \cite{WA,AW}

However, a problem with multiple fluctuation exchange is that each
additional fluctuation propagator gives rise to an additional energy
integration and will thus contribute a small phase space factor, anywhere
from $\omega ^{2}$ to $\omega $, depending on the critical momentum
dependence. As we will show below, such factors may be offset by inverse powers of the Fermi-liquid renormalization factor $Z^{-n}>>1$, provided the quasi-particle
weight factor $Z$ tends to zero, as the excitation energy $\omega$ or the
temperature $T$ tends to zero. As discussed in previous work by two of us, \cite{WA,AW} even if $Z\rightarrow 0$ in the limit $(\omega ,T)\rightarrow 0$ the
quasiparticle picture may still be applicable at non-zero $\omega ,T$, since
the quasiparticle width $\Gamma$ gets renormalized by a factor of $Z$,
which helps to keep it smaller than the excitation energy $\omega $, a
necessary condition for the existence of a quasiparticle peak in the
single-electron spectral function.

A careful identification of the effects of the quasiparticle critical behavior on the 
bosonic (AFM) propagator is at 
heart of our theory. This includes the determination of critical vertex
corrections of various types. The interplay of spin fluctuations and
fermionic excitations has also been considered in $1/N$ expansion by Abanov
and Chubukov \cite{abch} and Abanov, Chubukov, and Schmalian,\cite{Abanov2003} and a renormalization group formulation has been
given by Metlitski and Sachdev. \cite{metsach} However, these authors did
not consider the quasiparticle renormalization in the strong coupling regime.

The quantum critical point in heavy fermion compounds has often been
associated with a breakdown of the Kondo effect and therefore a breakdown of
the picture of heavy quasiparticles. \cite{Si,pc,SVS} In this approach, it
is assumed that at the critical point the energy scales of the Kondo effect
and the exchange interaction between the localized $f$-spins (in the absence
of the Kondo effect) are approximately equal. Some of these scenarios have
been developed enough to allow comparison with experimentally observed
critical exponents, in particular for CeCu$_{5.9}$Au$_{0.1}$ and for YbRh$%
_{2}$Si$_{2}$. \cite{CeCuAu,pep} However, experimentally the quasiparticle
mass does not appear to be drastically reduced (by orders of magnitude) when
the QCP is approached, as would be expected if the Kondo effect were to be
suppressed. We argue here that in the cases we consider, the Kondo effect, or
more precisely the heavy quasiparticle picture, remains intact. However, the
quasiparticles experience an AFM spin-exchange interaction responsible for
the ordering of their spins in the AFM state. In other words, we propose
that the ordered state is an itinerant heavy-quasiparticle SDW state, at
least in the neighborhood of the critical point. The resulting small ordered
magnetic moment is in agreement with observation. Roughly speaking, the
ordered moments provide a magnetic field acting on the Kondo ions. As long
as the $f$-electron Zeeman splitting caused by this field is small compared
to the Kondo temperature, the Kondo effect is only weakly suppressed.


In this report, we present a semi-phenomenological theory of the scaling
behavior near an AFM QCP. As discussed above, we show how spin-exchange energy
fluctuations may lead to a momentum-independent critical quasiparticle self
energy. The feedback of the critical quasiparticle properties (the $Z$%
-factor) into the spin and energy fluctuation spectrum leads to a
self-consistent equation for the quasiparticle self energy and effective
mass $m^{\ast }/m\propto Z^{-1}$. This allows a strong-coupling solution in
the form of a fractional power law $Z(\omega )\propto \omega ^{\eta }$. The
value of $\eta$ for different circumstances will be discussed in Sec.\ IV.
In Sec.\ V, we show that the dynamical structure factor satisfies $\omega /T$-scaling
within the quantum critical region of the phase diagram. 
The free energy obeys scaling characterized by fractional power laws; this is described in Sec.\ VI . In Sec. \ VII, we present an alternative derivation
of our results in the framework of the spin-fermion model. Comparison of our
theory with experimental data is discussed in Sec.\ VIII and we
summarize our findings  in Sec.\ IX.

The above scenario depends sensitively on the detailed nature of spin
fluctuations in a given system. For example, 3D AFM fluctuations do not lead
to true critical behavior, i.e. a Gaussian fluctuation theory is applicable, 
\cite{Hertz,Millis} provided the effective mass enhancement by critical
fluctuations is not too large. We argue below that in CCA there is a wide
region of 2D antiferromagnetic fluctuations, which gives rise to a
substantial enhancement of the effective mass and may drive the system into
a strong-coupling regime of 2D or 3D antiferromagnetic fluctuations. In YRS,
on the other hand, ferromagnetic fluctuations in the temperature regime $%
0.3K\lesssim T\lesssim 20K$ lead to a substantial enhancement of the
effective mass when the system crosses over into a 3D antiferromagnetic
fluctuation regime below $0.3K$.

\section{Critical quasiparticles}

Our starting point is a heavy Fermi liquid as in an Anderson lattice
model of correlated $f$-electrons (on-site interaction $U$) hybridizing with
conduction electrons. The energy scale of the heavy-fermion band is given by
the \textquotedblleft coherence temperature" $T_{coh}$, which is well above
the temperature regime for which \textquotedblleft non-Fermi liquid"
behavior is observed near a QCP. On top of the heavy-fermion quasiparticle
renormalization, the critical fluctuations cause a further renormalization
on which we focus here. The single-particle Green's functions may be
decomposed into a quasiparticle term and an incoherent contribution, $G(%
{\bf k},\omega )=ZG^{qp}+G^{inc}$, where the quasiparticle weight factor $%
Z$ determines the quasiparticle effective mass $m^*$ and is defined by $Z^{-1}=1-\partial {\rm Re}\Sigma (\omega )/\partial
\omega = m^{\ast}/m$. Here, $\Sigma(\omega)$ is the electron self energy, whose real and imaginary parts determine the quasiparticle properties. The quasiparticle Green's function is given by $G^{qp}({\bf k}%
,\omega )=[\omega -E_{k}-i\Gamma ]^{-1}$, with $E_{k}=(m/m^{\ast
})v_{F}(k-k_{F})$, where $v_{F}$ is the Fermi velocity of the heavy-fermion
band and the quasiparticle width is $\Gamma =Z\,{ {\rm Im}\Sigma (E_{k})}$.

The condition for the quasiparticle picture to be valid is $\Gamma <|E_{k}|$, which
is satisfied in the Fermi-liquid regime, ($\Gamma =cE_{k}^{2}\ll |E_{k}|$ in the limit 
$E_{k}\rightarrow 0$). Here, we argue that this quasiparticle stability
condition may be satisfied even in non-Fermi liquid situations. We extend
the usual quasiparticle picture by recognizing that the parameter $%
Z=m/m^{\ast }$, as defined above, depends on the energy scale, $Z=Z(\omega )$.
Since the
(retarded) self energy is an analytic function in the upper-half $\omega$ plane,
the real and imaginary parts of any nonanalytic term (in the lower half
plane) are locally connected. Even in a true non-Fermi-liquid phase with $%
\Sigma (\omega )\propto -i(i|\omega |)^{1 -\eta }$, $\eta <1$ so that $%
{ {\rm Im}\Sigma (\omega )\propto {\rm Re}\Sigma (\omega )\propto |\omega
|^{1-\eta }}$ and $Z\propto (|E_{k}|)^{\eta }$, one finds $\Gamma
/|E_{k}|=\tan (\frac{\pi }{2}\eta )<1$ for $0<\eta <1/2$. In this case,
although $Z=0$ at the Fermi surface, the spectral function for non-zero
excitation energy may be peaked sharply enough to separate a quasiparticle
contribution from the incoherent part. In Fig.\ 1, we show the spectral function ${\rm Im} \,G(\bf{k},\omega)$ for various choices of this ``non-Fermi-liquid exponent" $\eta$, demonstrating that there is still a well-defined quasiparticle peak.
\begin{figure}[ht]
\centering
\includegraphics[width=.46\textwidth, viewport=100 200 600
660,clip]{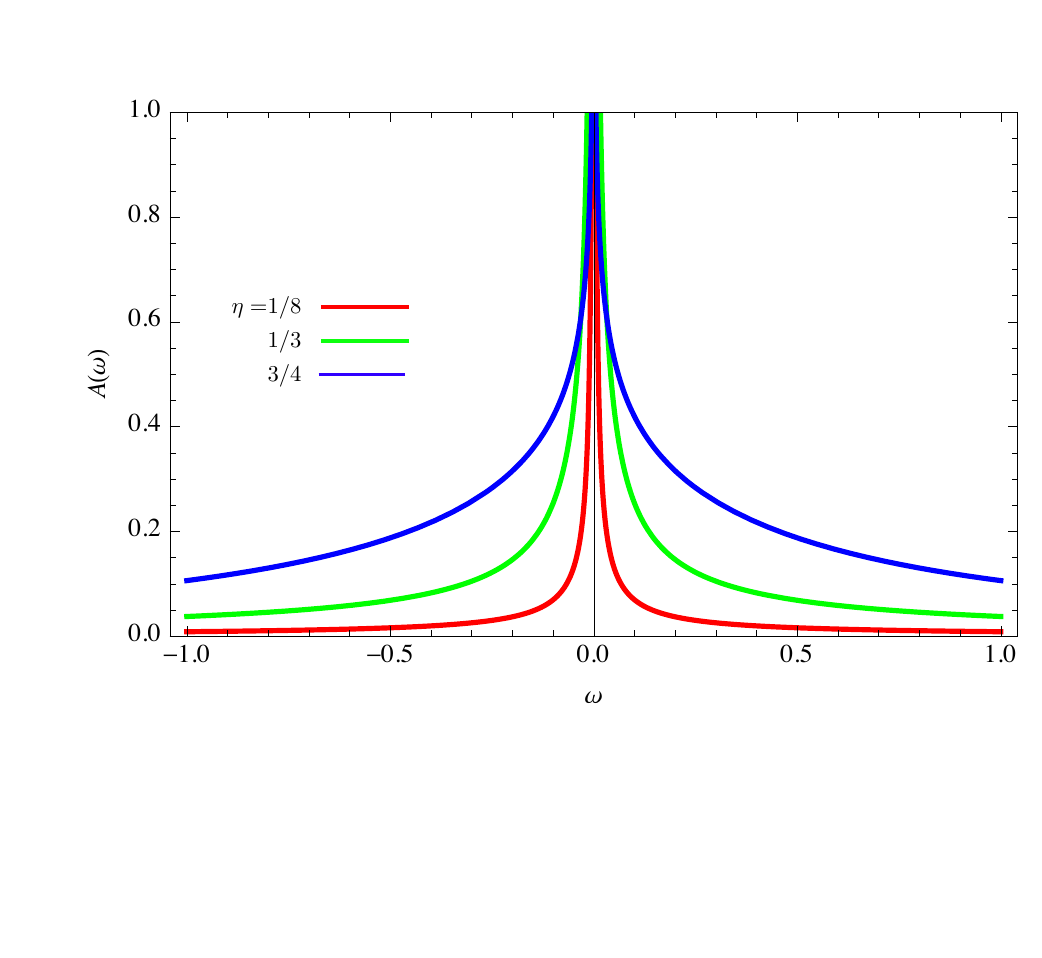}\vskip -1.5cm
\caption{Non-Fermi-liquid spectral function, for self energy $\Sigma (\omega )\propto -i(i|\omega |)^{1 -\eta }$, with $\eta =1/8, 1/3, 3/4$ (red, green blue, resp.)}\label{nflspectral}
\end{figure}
\section{Critical fluctuation spectrum}

We assume that in the paramagnetic phase of a metal close to an
antiferromagnetic quantum critical point, the self energy for the
single-particle Green's function is determined by the interaction with
magnetic fluctuations. We take the imaginary part of the renormalized
retarded dynamical spin susceptibility for wave vectors near the AFM
ordering wave vector ${\bf Q}$ to be of the form 
\begin{equation}
{\rm Im}\chi ({\bf q},\nu )=\frac{N_0(\nu\lambda_Q^2/v_F Q)}{[(r+{\bf q}^2\xi_0^2]^2 +(\nu\lambda_Q^2/v_F Q)^2},
\end{equation}
where $\bf{q}$ is measured from the ordering wave vector $\bf{Q}$.
Here, $N_{0}$ is the bare density of states at the Fermi surface, $v_{F}$ is
the bare Fermi velocity, and $\xi _{0}\simeq k_{F}^{-1}$ is the microscopic
AFM correlation length. The control parameter $r$ is a function of both the
tuning field and the temperature. The factor $\lambda_Q$ is the low-energy
spin vertex at $\bf{q}=\bf{Q}$. A microscopic derivation of its
behavior is in preparation.\cite{pw-cmv} We expect it to have singular
behavior \textit{e.g.} $\lambda_Q \sim \omega^{-\phi}$, similar to that of
the ${\bf q} = 0$ vertex for which $\phi =\eta$,  the $Z$-factor exponent, by virtue of a Ward
identity. In what follows, we use the results of Ref.\ \onlinecite{pw-cmv} and choose $\phi =\eta$ also for $\lambda_Q$. In fact, we show in this paper that
this choice enables excellent agreement with experiment.

At $T=0$, the control parameter $r$ vanishes at the QCP as a fractional power 
$r\propto |r_0|^{2\nu}$, where $r_{0}\propto(H-H_{c})$ or $r_{0}\propto(P-P_c)$, for magnetic field or pressure tuning, and $\nu$ is the correlation length exponent. In Appendix
A, we show that $\nu = 1/(2 +z\eta)$ [Eq.\ (A1)], where $z$ is the dynamical
exponent, determined in  Sec.\ V. The non-Fermi liquid exponent $\eta$ is self-consistently determined in Sec.\ IV.

We now define an energy-fluctuation propagator $\chi _{E}({\pmb\kappa },\nu )
$ by combining two spin fluctuation propagators, in the form 
\begin{align}
{\rm Im}\,\chi _{E}({\pmb\kappa },\nu )=& \sum_{\bf{q}_{1},\nu
_{1}}G_{k+q_{1}}G_{k+q_{1}-\kappa }{\rm Im}\chi (\bf{q}_{1},\nu _{1})
\\
\times & {\rm Im}\chi (\bf{q}_{1}-{\pmb\kappa },\nu _{1}-\nu )[b(\nu_{1}-\nu )-b(\nu _{1})],  \notag
\end{align}%
where $b(\nu )$ is the Bose function. The Green's functions $%
G_{k+q_{1}},G_{k+q_{1}-\kappa }$ are off-shell for most values of the
momentum $\bf{q}_{1}$\ and each may be replaced by $1/\epsilon _{F}$.
Performing the $d$-dimensional momentum integration by Fourier transform, we get

\begin{align}
{\rm Im}\chi _{E}({\pmb\kappa },\nu )\approx & N_{0}^{3}\lambda_Q^{-2}(%
\frac{\nu\lambda_Q^2 }{v_F Q})^{d-1/2} \\
& \times \frac{1}{[\frac{\nu\lambda_Q^2 }{v_F Q}+\kappa ^{2}\xi
_{0}^{2}+r]^{(d+1)/2}},  \notag
\end{align}%
%

\section{Self-consistent determination of Quasiparticle self energy}

Now we set up a self-consistent determination of the quasiparticle self
energy via the leading term in a skeleton graph expansion in terms of the
boson propagator $\chi _{E}$. The imaginary part of the self energy is given
by

\begin{align}
{ {\rm Im}\Sigma (k,\omega )}& {\rm=-\lambda _{E}^{2}\int \frac{d\nu }{%
\pi }\sum_{\bf{q}}{\rm Im}G(\bf{k}+\bf{q},\omega +\nu )} 
\notag \\
& \times {\rm Im}\chi _{E}({\bf q},\nu )[b(\nu )+f(\omega -\nu )] 
\notag \\
& \approx v_F Q Z^{-2}(\omega\lambda_{Q}^2 /v_F Q)^{d-1/2}  \notag \\
& \propto |\omega |^{d-1/2-\eta(2d+1)},
\end{align}
The interaction vertex $\lambda _{E} = \lambda_Q^2 \lambda_v$, where $%
\lambda_v$ is $\propto Z^{-1}$, as it arises through a Ward identity
connected to energy conservation. We used $\lambda_{Q}\propto Z^{-1}$, as
discussed above, below Eq.\ (1). The Fermi and Bose functions $f(\omega ),b(\omega )
$ confine the $\nu $-integration at low $T$ to the interval $[0,\omega ]$.
In Eq.\ (4), we used the power law $Z(\omega )\propto |\omega |^{\eta } $ .
The scale-dependent contribution to ${\rm Re}\,\Sigma (\omega )$ follows
from analyticity as ${\rm Re}\,\Sigma (\omega )\propto (\omega /v_F
Q)^{d-1/2} Z^{-(2d+1)})$ . This leads to the self-consistency relation for $%
Z(\omega )$

\begin{align}
Z^{-1}(\omega ) &=1-\partial {\rm Re}\Sigma (\omega )/\partial \omega 
\notag \\
& \approx 1+Z^{-2d-1}(\omega /v_FQ)^{d-3/2}
\end{align}

We now explore the consequences of the scale-dependent $Z$. In general, the $%
\omega$ and $T$ dependence of $Z$ is obtained by substituting $\sqrt{\omega
^{2}+a^2 T^{2}}$ for $\omega $, where $a$ is a constant of order unity. For
frequencies less than the temperature, we may replace $\omega $ by $T$. As
long as $Z^{-2d-1}(T /v_FQ)^{d-3/2}\ll 1$ for any $T$, the system will be in
the Gaussian fluctuation regime all the way down to the critical point. If,
however, the initial value of $Z^{-1}(T)$, when one enters the AFM
fluctuation regime, is sufficiently large, such that $Z^{-2d-1}(T
/v_FQ)^{d-3/2}\gg 1$, a new regime is accessed, which is of a
strong-coupling nature. We find the characteristics of this new regime
within the present approximation by solving the self-consistent Eq.\ (5), to
get 
\begin{equation}
Z(T)\propto (T/v_FQ)^{\eta},
\end{equation}%
where the exponent $\eta$ is found to be 
\begin{equation}
\eta =(2d-3)/4d.
\eq

In the case of only AFM fluctuations in a clean system, it is difficult to
satisfy the strong-coupling condition of sufficiently large $Z^{-1}(T)$ .
However, if on the initial approach to the critical point, fluctuations
dominate that lead to a growing $Z^{-1}(T)$ with decreasing $T$, the
condition may be met at some point. The precise crossover point is
determined by the crossover of these precursor fluctuations to the critical
AFM fluctuations and by the condition above that leads to Eq.\ (6). As
discussed in \cite{WA,AW} impurity scattering helps to enhance the effect of
AFM fluctuations on $Z(T)$ . In addition, there are clear indications in the
data on YbRh$_{2}$Si$_{2}$ of 3D FM fluctuations. \cite{FM} In that case,
one finds $Z^{-1}(T)\propto \ln (T_{0}/T)$, so that indeed $Z^{-1}$ grows as 
$T\rightarrow 0$ .

\section{Critical exponents and dynamical scaling}

The critical behavior of the spin-excitation spectrum as discussed above,
see Eq.\ (1), is given by (employing units $(v_FQ,\xi _{0}^{-1})$ for $%
(\omega ,q)$)

\begin{equation}
{\rm Im}\chi ({\bf  q},\omega )\propto \frac{\omega^{1-2\eta}}{[\xi
^{-2}+q^{2}]^{2}+(\omega^{1-2\eta})^{2}},
\end{equation}

where 

\bq
\xi = r^{-1/2}.
\eq
By equating the terms $q^{2}$ and $\omega ^{1-2\eta}$ in the denominator, we
find the dynamical critical exponent as 
\bq
z=2/(1-2\eta)=4d/3.
\eq

Thus, in the critical region, the leading temperature dependence of the correlation length is given by $\xi (H_{c},T)\sim T^{-1/z}$ and using Eq.\  (A1), we find the
correlation length exponent  $\nu =
3/(3+2d)$. The boundary of the critical region - the ``critical cone" - in the $(H,T)$
phase diagram is found from $\xi (H,T=0)=\xi (H_{c},T)$ as $T\sim
|H-H_{c}|^{z\nu }$.

Then $1/\xi (r_{0},T)$ has the form 
\begin{equation}
1/\xi (r_{0},T)=T^{1/z}g(r_{0}T^{-1/\nu z}),
\end{equation}
where $g(x)\approx 1+x^{\nu }$ . The $(\omega ,T)$ dependence of $Z$ may be
accounted for with the form $Z(\omega ,T)\propto T^{\eta}\zeta (\omega /T)$,
where $\zeta (x) \propto (x^2 +a^2)^{\eta/2}$.

Then 
\begin{equation}
{\rm Im}\chi ({\bf q},\omega )\propto T^{-2/z}\frac{(\omega /T\zeta
^{2})}{[g^{2}+T^{-2/z}q^{2}]^{2}+(\omega /T\zeta ^{2})^{2}},
\end{equation}%
which, with the use of Eq.\ (8), shows the following general scaling relation

\begin{equation}
{\rm Im}\chi ({\bf q},\omega )\propto T^{-2/z}\Phi (\frac{\omega }{T}%
,q\xi ;r_{0}T^{1/z\nu }).
\end{equation}

Inside the critical cone, we may set $r_{0}=0$ so that $g=1$\ and $\xi
^{-1}\propto T^{1/z}$. Defining $x=\omega \xi ^{z}=\omega /T$, and $y=q\xi $%
, we find the scaling form

\begin{eqnarray}
{\rm Im}\chi ({\bf q},\omega ) &\sim &\xi ^{2}\Phi (x,y), \\
\Phi (x,y) &=&\frac{x/\zeta ^{2}(x)}{(1+c_{q}y^{2})^{2}+(x/\zeta ^{2}(x))^{2}%
},
\end{eqnarray}%
where $c_{q}$ is a constant.

At the ordering wave vector, $y=0$ so  the spin excitation
spectrum obeys $\omega /T$-scaling  inside the critical cone:
\begin{equation}
{\rm Im}\chi ({\bf Q},\omega )\sim T^{-2/z}\frac{\omega /T\zeta
^{2}(\omega /T)}{1+(\omega /T\zeta ^{2}(\omega /T))^{2}},
\end{equation}

A comparison of this scaling form with neutron scattering data on CeCuAu is
shown in Fig.\ 2, where we used $d=2$, so that $z=8/3$ and $\eta = 1/8$.
\begin{figure}[h]
\centering
\includegraphics[width=.44\textwidth, viewport=50 40 660 480,clip]{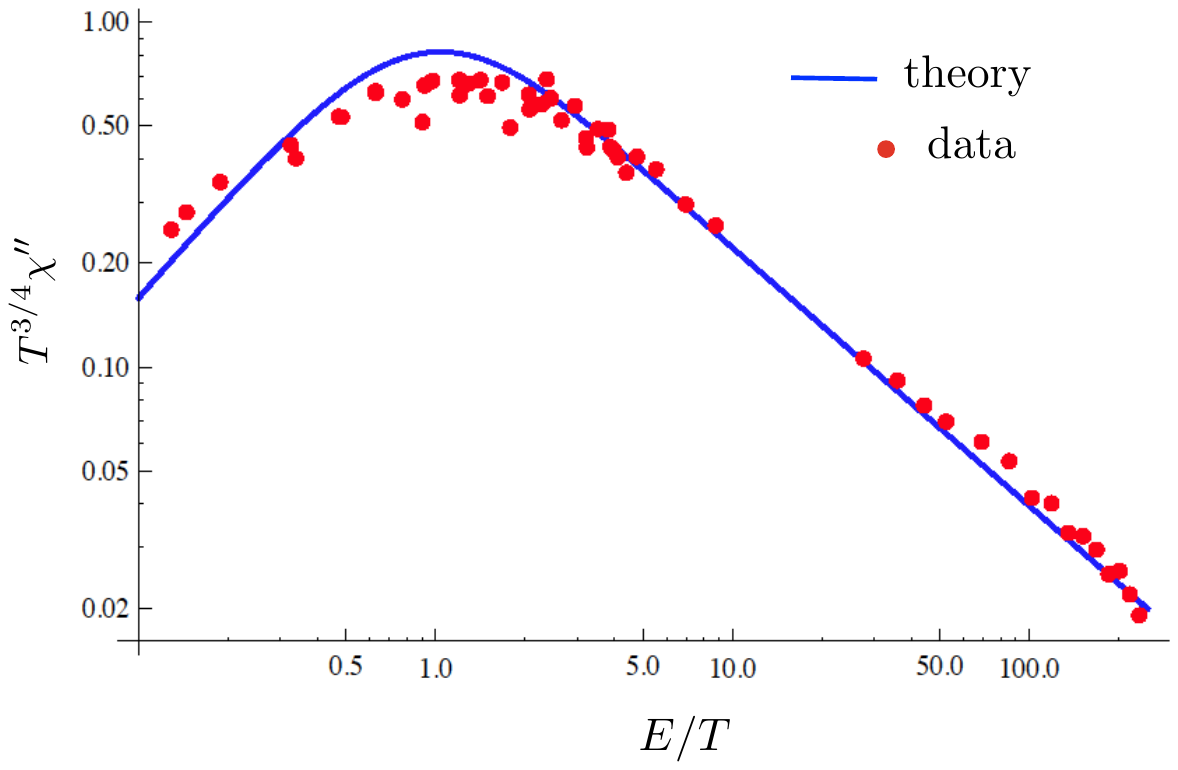} \vskip -.1cm
\caption{Inelastic neutron scattering: Comparison of theory Eq.\ (16) and
experimental data \protect\cite{schro} for CeCu$_{6-x}$Au$_{x}$ at the
critical concentration $x=0.1$}
\label{ins}
\end{figure}

\section{Free energy}

The critical part of the free energy density may be derived from the
expression for the entropy density in terms of the self-energy \cite{chu}:

\begin{equation}
\frac{S}{V}=\frac{1}{2\pi }N(0)\int \frac{d\omega \omega }{T^{2}\cosh ^{2}%
\frac{\omega }{2T}}[\omega -{\rm{Re}\Sigma (\omega )]}
\end{equation}%
Substituting the critical self energy found above and integrating over
temperature, we find the scaling form of the free energy density in the case
of $d$-dimensional spin fluctuations ($d=2,3$) in a $3D$ metal

\begin{equation}
f(H,T)= \xi^{-(2d+1)}\Phi _{f}(r_{0}\xi ^{1/\nu },T\xi ^{z})
\end{equation}

The correlation volume $V_{c}\sim \xi ^{(2d+1)}$ is understood as follows:
The underlying critical degrees of freedom in the very low temperature
Anderson lattice picture are the heavy fermonic quasiparticles described by
the propagator $G(k,\omega )^{-1}=\omega -vk-\Sigma (\omega )$ with $\Sigma
(\omega )\propto \omega ^{1-\eta }$. Therefore, their dynamical exponent $%
z_{f}=1/(1-\eta )$ and their dimensionality is $d_{f}=1$ \cite{senthil}. The
entropy of the system is determined by hyperscaling for the fermions: $%
S\propto T^{d_{f}/z_{f}}$. Therefore, since $\xi \propto T^{-1/z}$, where $%
z=4d/3$, the free energy density $f\propto T^{1+d_{f}/z_{f}}\sim \xi ^{2d+1}$%
. Here, $z$ and $d$ are the bosonic exponents discussed above.

The specific heat coefficient follows as

\begin{equation}
C/T \propto 
\begin{cases}
T^{(2d+1-2z)/z}, & \text{critical regime} \\ 
r_{0}^{\nu (2d+1-2z)}, & \text{Fermi liquid regime.}%
\end{cases}%
\end{equation}

For the critical part of the magnetization we find

\begin{equation}
M-M(H_{c},0)\propto 
\begin{cases}
-T, & \text{critical regime} \\ 
-r_{0}^{\nu (2d+1)-1}, & \text{Fermi liquid regime.}%
\end{cases}%
\end{equation}

The susceptibility has a critical part 
\begin{equation}
\chi -\chi (H_{c},0)\propto 
\begin{cases}
-T^{1-1/\nu z}, & \text{critical regime} \\ 
-r_{0}^{\nu (2d+1)-2}, & \text{Fermi liquid regime.}%
\end{cases}%
\end{equation}

Using $\partial M/\partial T=-{\rm const}$, we find the Gr\"{u}neisen ratio
in the critical regime: 
\begin{equation}
\Gamma _{G}=-\frac{\partial M/\partial T}{C}\propto \frac{1}{C}\propto
T^{(z-2d-1)/z}
\end{equation}%
while in the Fermi liquid regime, we have the universal result

\begin{equation}
\Gamma _{G}=-\frac{G_{r}}{H-H_{c}}\text{ , \ \ \ }G_{r}=-(z-\frac{2d+3}{3}%
)\nu
\end{equation}

The critical scaling of transport properties is obtained by observing
that the quasiparticle relaxation rate scales as $\Gamma \propto \xi
^{-z}\Phi _{\Gamma }(r_{0}\xi ^{1/\nu },T\xi ^{z})$. Then the resistivity
behaves as

\begin{align}
\rho -\rho (H_{c},0)&\propto \frac{m^{\ast }}{m}\Gamma \\
&\propto 
\begin{cases}
T^{(z+2)/2z}, & \text{critical regime} \\ 
-|r_{0}|^{(2-3z)\nu /2}T^{2}, & \text{Fermi liquid regime.}%
\end{cases}
\notag
\end{align}%
For the thermopower, $S$ we get, using $S\propto m^{\ast }T$, $S\propto C$
in both the critical regime and the Fermi liquid regime.

\section{Scaling properties within the spin-fermion model}

The results of the phenomenological theory, presented in the previous
sections, can alternatively be derived within an approach based upon
the spin-fermion model\cite{Abanov2003,metsach}, assuming fermions $\psi$ to couple
to a spin-1 boson field $\boldsymbol{\phi}$ describing collective antiferromagnetic spin
fluctuations according to 
\begin{equation}
S_{int}=g\int d^{d}x\psi^{\dagger}\bf{\boldsymbol{\sigma}}\psi\bf{%
\bf{\cdot\boldsymbol{\phi}}}\, .
\end{equation}
As mentioned in Sec.\ I, the role of critical fluctuations for the fermionic
spectrum was traditionally studied at and near hot spots or lines of the
Fermi surface, where $\varepsilon_{\bf{k}_{F}}=\varepsilon_{\bf{k}%
_{F}\bf{+Q}}$ with magnetic ordering vector $\bf{Q}$. For example,
the self-energy in one-loop approximation at hot spots behaves\cite{Abanov2003} as $%
\Sigma_{h}\left(\omega\right)\propto g^{2}i\omega\left|\omega\right|^{\frac{%
d-3}{2}}$ for a $d$-dimensional system. Note that in this Section, $\omega$ is a Matsubara frequency. Thus for $d\leq3$ one obtains
deviations from Fermi-liquid behavior,\cite{sc} $\Sigma_{FL}\left(\omega\right)%
\propto i\omega$, for hot momenta. Up to order $g^{2}$, Fermi-liquid behavior
occurs for generic, cold quasiparticles. However Hartnoll et al.\cite%
{hart} demonstrated that higher order processes, involving fermions
which effectively couple via off-shell intermediate states to $\bf{\phi}%
^{2}$, affect the cold regions as well, albeit with less strong
singularities in the single particle excitation spectrum. Generalizing their
findings to arbitrary dimension, one finds a self-energy contribution on the
cold parts of the Fermi surface $\Sigma_{c}\left(\omega\right)\propto
g^{4}i\omega\left|\omega\right|^{d-\frac{3}{2}}$, as we previously obtained in Eq.\ (4). Even though this behavior
is subleading with respect to Fermi liquid behavior for $d>\frac{3}{2}$, it
constitutes a singular correction. It is at the heart of our theory to show how 
this singular correction may be boosted within a self-consistent approach.

To extract this important physics, we \textquotedblleft
patch\textquotedblright\ the Fermi surface into hot and cold regions $(h,c)$ and
start from a bare action 
\begin{eqnarray}
S &=&\sum_{j=h,c}\int [\psi _{j}^{\dagger }\left( i\omega -\varepsilon _{\bf k}^j
\right) \psi _{j}+g\psi _{j}^{\dagger}\boldsymbol{\sigma }\psi _{j}\cdot \boldsymbol{\phi } +\lambda \psi^{\dag}_j\psi_j\boldsymbol{\phi}\cdot\boldsymbol{\phi}]  \notag \\
&&+\frac{1}{2}\int (r_{0}+q^{2})\boldsymbol{\phi}\cdot \boldsymbol{\phi}.
\end{eqnarray}%
 Here, $\lambda$ is the coupling of quasiparticles to a pair of bosons via intermediate off-shell fermonic states, \cite{hart} as mentioned above.  If we couple an external
magnetic field to the electron spin, $\int \bf{h}\cdot \psi ^{\dagger}\boldsymbol{\sigma}\psi$, we may shift $\boldsymbol{\phi}^{\prime}=\boldsymbol{\phi}-\bf{h}/{\it g}$ and obtain the relation $\chi({\bf q},\omega) =D\left( \bf{q},\omega \right) -r_{0}^{-1}$, which connects the
spin-susceptibility $\chi \left( \bf{q},\omega \right)$ and the
propagator $D\left( \bf{q},\omega \right)$ of the bosons. Let $\Pi
\left( \bf{q},\omega \right) $ be the full bosonic self energy. Then
\begin{equation*}
D\left( \bf{q},\omega \right) =\frac{1}{r_0+q^2 -\Pi \left( \bf{q},\omega
\right) }.
\end{equation*}%
Close to the critical point at the ordering vector ($q=0$), $r_0\simeq \Pi \left({\bf Q}, 0\right)$.
This model is valid up to the band width, $W$.

We now set up a matching procedure by first integrating out all states down to an energy $\Lambda <W$. This introduces boson and (hot and cold) fermion self energies into the bare action of Eq.\ (26) as well as renormalized  couplings $\Gamma$ among the fields. The effective low-energy action is

\begin{align}
S_{low} &=&\sum_{j=h,c}\int \psi _j^{\dagger} \left[
i\omega -vk-\Sigma _j^> ( k,\omega)
+\Gamma _{j,g}^>\bf{\sigma}\cdot \bf{\phi}\right]\psi_j \nonumber
 \\
&+&\frac{1}{2}\int \left[ r_{0}+q^{2}-\Pi ^>( q,\omega) \right]
 {\bm \phi}^2+\int \Gamma _{\lambda}^>\psi _c^{\dagger}\psi _c{\bm \phi}^2. 
\end{align}%
We have taken the bare Fermi velocities of both hot and cold electrons to be equal, for simplicity.
The relevant bare couplings are $g$ and $\lambda \simeq g^{2}/E_{F}$. 
Here $\Sigma _{j}^{>}\left( k,\omega \right) $ etc. refer to scattering
processes that involve energies above $\Lambda$. The behavior in the high-energy region is governed by the bare action [Eq.\ (26)], while the behavior for energies below $\Lambda$ is governed by $S_{low}$. In what follows, we shall match the low and high energy sectors at $\Lambda$.

We parameterize
the self energy and vertex functions as: 

\begin{eqnarray}
\Sigma _{j}^{>}\left( k,\omega \right)  &=&\left( 1-Y_{j,\omega
}\right) i\omega +v\left( Y_{j,k}-1\right) k  \notag \\
\Gamma _{j,g}^{>}\left( k,q,\omega ,\Omega \right)  &=&gY_{j,g}  \notag
\\
\Gamma _{\lambda }^{>}\left( k,q,\omega ,\Omega \right)  &=&\lambda
Y_{c,g}^{2}Y_{\lambda }
\end{eqnarray}%
The $Y_j$ are all functions of $\Lambda$. 

Introducing quasiparticle operators $\psi
_{j,r}=\psi _jY_{j,\omega }^{1/2}$,  we arrive at the low-energy
quasiparticle action 
\begin{eqnarray}
S_r &=&\sum_{j=h,c}\int \left[ \psi _{j,r}^\dagger( i\omega
-v_j^*k) \psi _{j,r}+g_j^*\int \psi _{j,r}^\dagger%
\bm{\sigma}\psi _{j,r}\cdot{\bm \phi}\right]   \notag \\
&&+\frac{1}{2}\int\left[ r_0-\Pi^>(q,0)+q^2\right] {\bm \phi}^2\notag\\&&%
+\lambda ^*\int \psi _{c,r}^\dagger\psi _{c,r}{\bm \phi}^2.%
\end{eqnarray}
that is governed by renormalized velocities and coupling constants: $%
v_{j}^{\ast }=vY_{j,k}/Y_{j,\omega }$, $g_{j}^{\ast
}=gY_{j,g}/Y_{j,\omega }$, and $\lambda ^{\ast }=\lambda
Y_{c,g}^{2}Y_{\lambda }/Y_{c,\omega }$. In the following, we shall label the self-energies of the $\psi_r$ field by a subscript ``$qp$".

There are no singular corrections to the boson self energy; its non-singular frequency and momentum behavior can be  absorbed into the existing frequency and momentum dependence and it is unnecessary to introduce a renormalized ${\bm \phi}$ operator. Finally, the boson self energy $\Pi_{qp}(q,\omega)$ is determined perturbatively, see below.

The renormalization of the coupling
constant $\lambda $ is a consequence of the composite nature of its
coupling and reflects the fact that the coupling to energy-density
fluctuations is affected by renormalizations $Y_{g}$ of the coupling
constant $g$.

The key idea is to develop a perturbation theory in the low-energy sector in terms of the
renormalized coupling constants. To this end, we take advantage of the fact
that the momentum dependence of the self energy is weak, so that  $%
Y_{j,k}\approx 1$. In addition we use the small-$q$ Ward identity $%
Y_{\lambda }=Y_{c,\omega }$, which reflects energy conservation.

Making contact with the earlier Sections, we introduce 
 $Z_j^{-1}=Y_{j,\omega }$ for the quasiparticle weights. $Z_c$ is to be identified with the  $Z$ of Secs. I-IV. We  obtain $v_{c}^{\ast }=vZ_c$, $v_{h}^{\ast }=vZ_{h}$, $g_c^*=Z_cY_{c,g}$, $g_h^{\ast }=gZ_{h}Y_{h,g}$, and $\lambda ^{\ast }=\lambda Y_{c,g}^{2}$. This
leaves us with three unknown renormalization factors. At a quantum critical
point one expects power law solutions of the kind $Y_{h,g}\propto Y_{c,g}\propto \left\vert
\omega \right\vert ^{-\phi }$(we assume the two $Y_j$ to be governed by the same exponent $\phi$), $Z_{h}\propto \left\vert \omega \right\vert
^{\eta _{h}}$, and $Z_c\propto \left\vert \omega \right\vert ^{\eta_c }$. These
three exponents lead to corresponding dynamic scaling exponents of the
critical degrees of freedom: spin-fluctuations, hot quasiparticles, and cold
quasiparticles. In the case of the fermionic degrees of 
freedom, we obtain dynamic scaling exponents $z_{f,c}=\frac{1}{1-\eta_c }$
for cold and $z_{f,h}=\frac{1}{1-\eta _{h}}$ for hot portions of the Fermi
surface.
 For the bosonic self energy due to a quasiparticle loop, it follows from low-energy perturbation
theory that 
\begin{equation*}
\Pi _{qp}(q,\omega) \approx \Pi _{qp}(
Q,0)-\gamma |{\omega}|
\end{equation*}%
with the coefficient of the Landau damping  $\gamma =(g_h^*/v_h^*)^2/Q = (gY_{h,g}/v)^2/Q$. \cite{Pi}

Inserting this result
into the bosonic propagator, we find at the critical point the dynamic bosonic
scaling exponent
\begin{equation}
z=\frac{2}{1-2\phi }.
\end{equation}%
As pointed out earlier, an anomalous dynamic critical exponent, as seen in
the experiments by Schr\"{o}der et al. \cite{schro} requires a nontrivial
exponent $\phi $ in the vertex renormalization. A straightforward
perturbation theory (at $T=0$) with respect to $\lambda ^{\ast }$ yields the
self energy of renormalized quasiparticles on the cold parts of the Fermi surface:
\bq
\Sigma _{qp,c}\left( i\omega \right) \propto \frac{\lambda ^{* 2 }}{v_c^*} \gamma^{d-5/2} {\rm%
sign}\left( \omega \right) \left\vert \omega \right\vert ^{\frac{2d-1}{2}},  \label{pertcold}
\eq%
while perturbation theory with respect to $g_{h}^{\ast }$ yields for hot
carriers 
\begin{equation}
\Sigma _{qp,h}\left( i\omega \right) \propto \frac{g_h^{*2}}{v_h^*} \gamma^{d/2-3/2}  {\rm%
sign}\left( \omega \right) \left\vert \omega \right\vert ^{\frac{d-1}{2}}\, .
\end{equation}

Thus far, high and low energy processes have been treated rather
differently. Yet, they meet at the scale $\Lambda$. The matching 
is realized as follows: For each sector (hot,cold), the Green's function $G$ generated by the action $S_{low}$ 
is
matched to $G_{qp}$, the one generated by the low-energy ``quasiparticle" action  $S_r$ of Eq.\ (29). 
We find
\bq
G (\omega = \Lambda) = Z_s((\omega = \Lambda)G_{qp}(\omega = \Lambda),
\eq
where $1/Z_s(\omega)= 1 - \partial\Sigma^>(\omega)/\partial\omega$.

In a genuine strong-coupling regime, we expect that the quantum critical
behavior is not confined to low energies but extends to high energies. Then
the scale $\Lambda$ is not a physically motivated crossover
scale, but rather an arbitrary intermediate scale. Therefore, we can generalize Eq.\ (33) and request that for arbitrary $\omega$, 
\bq
\Sigma_{qp}(\omega) = Z_s(\omega)\Sigma^>(\omega),
\eq
where the LHS is given by the perturbative results of  Eqs.\ (31,32).  In addition, the matching yields that $Z_s$ is identical to $Z=1/Y_{\omega}$ of earlier paragraphs.  Knowing  the power-law behavior of $\Sigma_{qp}$, we may look for power-law solutions $Z_j \propto \omega^{\eta_j}$, with the result that\begin{align}
\eta_c&=d-\frac{3}{2}+\phi \left(1-2d \right),  \\
\eta _{h}&=\frac{3-d}{2} +\phi \left(d-1\right).
\end{align}%
Low-order perturbation theory within the spin-fermion model yields $\phi =0$
for $d>2$, {\em i.e.} there are no singular renormalizations of the fermion-boson coupling $g$ and correspondingly no changes in the dynamic scaling exponent $z$, Eq.\ (30)
from its mean-field value $z=2$. In this limit, we obtain for hot carriers
that $\eta _{h}\left( \phi =0\right) =\frac{3-d}{2}$.
Most interestingly, we do obtain a non-trivial result for the
anomalous exponent of the cold carriers $\eta \left( \phi =0\right) =d-3/2$. Thus, even at the lowest level the self-consistent perturbation theory presented here
yields genuine non-Fermi liquid behavior on the entire Fermi surface.

Our phenomenological framework developed here allows furthermore to go
beyond the spin-fermion model to include effects of higher order perturbation theory or due to additional
interaction channels. For example, as shown recently, \cite{CW} the actual derivation of the $q=0$
Ward identity requires the resummation of an infinite class of diagrams, including, in particular, 
diagram structures of the Azlamazov-Larkin type. \cite{AL} Recently, new Ward identities for the spin vertex that are valid at any wave vector ${\bf q}$ have been discovered, \cite{pw-cmv} which relate the spin vertex to the effective mass enhancement. As discussed above, we can explore the implications of this result, that the Ward identity for the spin-vertex at $%
{\bf q}=0$ carries over to the vertex at momentum transfer ${\bf q}={\bf Q}$, which means $Y_{h,g}=$ $Z_c^{-1}$. This relation immediately implies $\phi =\eta_c$, which turns Eq.\ (35) into a self-consistent equation for $\eta_c$ \
 and yields 
\begin{eqnarray}
\eta _{h} &=&\frac{3+d}{4d}\notag \\
\eta_c &=&\frac{2d-3}{4d} . 
\end{eqnarray}%
as was obtained in the phenomenological theory presented above, in Eq.\ (7), including
the dynamic scaling exponent $z$ obtained earlier in Eq.\ (10). In fact, $(Z_c, \; \eta_c)$ should be identified with $(Z,\eta)$ of previous sections I-IV

\section{Comparison with experiment}

The above results, in the case of 3D fluctuations, are identical to the ones
obtained by two of us previously, \cite{WA,AW} where impurity scattering was
invoked to give rise to a critical, weakly momentum-dependent self energy.
In particular, the $Z$-exponent $\eta =1/4$ found there, and the ensuing
critical indices $z=4,\nu =1/3$ are the same as the ones found in the
present work for the clean system. The excellent agreement of the theory 
\cite{WA,AW} with the experimental data on YbRh$_{2}$Si$_{2}$ (YRS) in the regime
close to the QCP therefore applies to the present theory as well. In
contrast to that earlier work, the present results do not depend on the
impurity concentration. Indeed, in experiment, the critical parts of {\it
e.g.} the specific heat do not show a dependence on the impurity content of
the sample.

We turn to a different case, CeCu$_{6-x}$Au$_{x}$ (CCA), for which a QCP has been
found at the concentration $x=0.1$ at ambient pressure in the absence of a
magnetic field and at slightly different concentrations at a critical
pressure and/or an applied critical field. As suggested by the neutron
scattering data, the magnetic fluctuations there appear to be
quasi-two-dimensional. We therefore compare our results for $d=2$ with the
available data. In doing this we keep in mind that generically in the case
of quasi-two-dimensional spin fluctuations in a three-dimensional metal, the
hot regions on the Fermi surface (the regions where both partners of a
particle-hole excitation at $q=Q$ are near the Fermi surface ($\epsilon _{{\bf k}}\approx ~\epsilon _{\bf{k+Q}}\approx 0$) occupy a finite
fraction of the Fermi surface (the scenario of quasi-two-dimensional
fluctuations in the 3D metal CCA was first proposed by Rosch {\it et al},\cite{rosch} in the framework of Hertz-Millis theory).
We assume this
fraction to be sufficiently small, such that over a wide intermediate
temperature range, quasiparticles on the cold parts of the Fermi surface
dominate. However, below some crossover temperature the critical behavior in
that case will be governed by the hot quasiparticles. 
 In Fig.\ 2, we have already compared theory and experiment for the dynamical
spin susceptibility. In Fig.\ 3, we show the specific heat data \cite%
{physicab} in comparison with $C/T\propto T^{-1/8}$ as obtained above in
Eq.\ (19)

\begin{figure}[ht]
\centering
\includegraphics[totalheight=.3\textheight, viewport=120 80 950
600,clip]{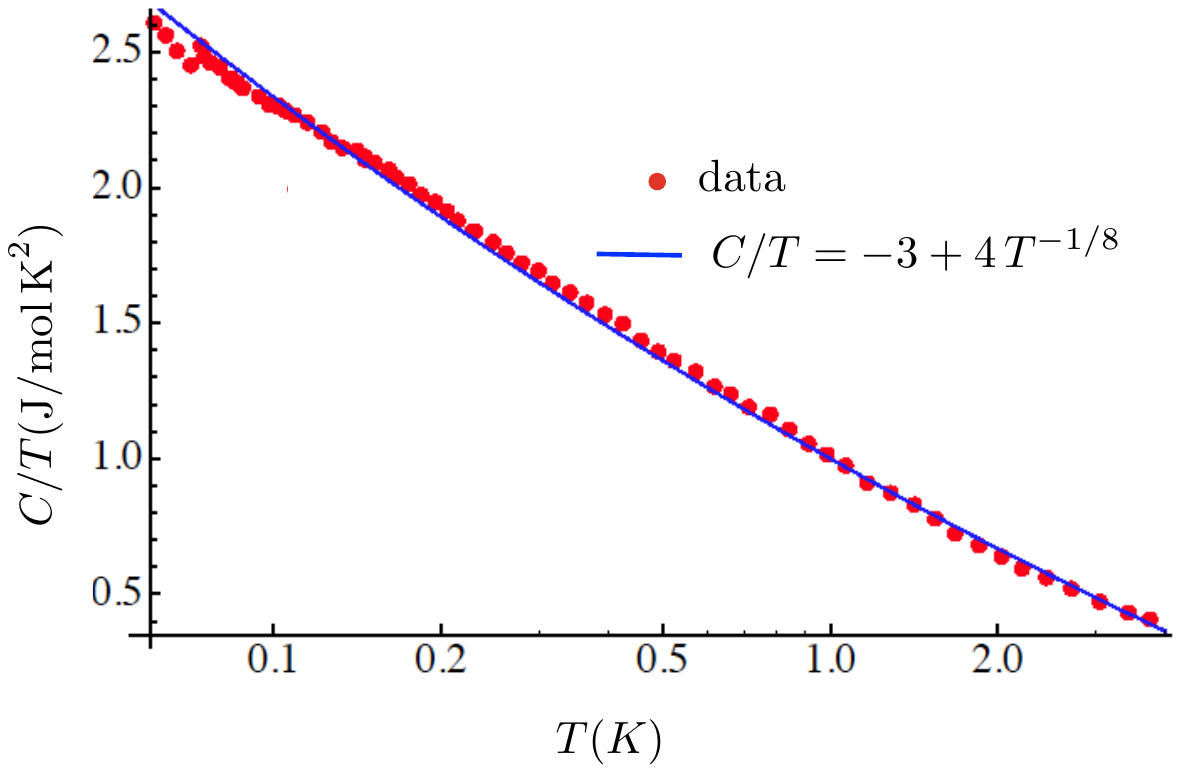}\vskip -1.2 cm
\caption{Specific heat: Comparison of theory Eq.\ (19) and experimental data 
\protect\cite{physicab} for CeCu$_{6-x}$Au$_x$ at the critical concentration 
$x=0.1$}
\label{C/T}
\end{figure}

The resistivity result $\rho (T)-\rho (0)\propto T^{7/8}$ of Eq.\ (24) is fitted to the
data in Fig.\ 4.
\begin{figure}[h]
\centering
\includegraphics[totalheight=.42\textheight, viewport=80 00 950
700,clip]{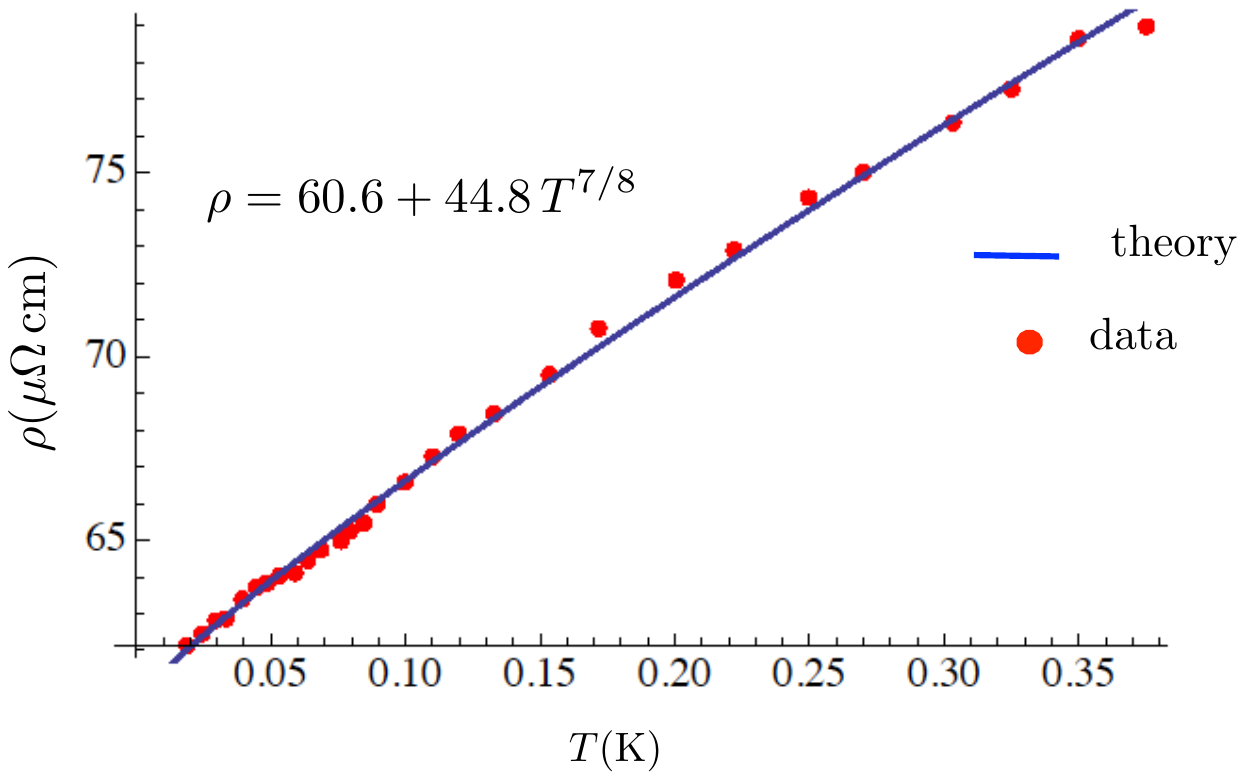} 
\caption{Resistivity: Comparison of theory Eq.\ (24) and experimental data 
\protect\cite{HvL2} for CeCu$_{6-x}$Au$_x$ at the critical concentration $x=0.1$}
\label{res}
\end{figure}

Our prediction for the uniform magnetization is $M(T)=M(0)-aT+bT^{2}$, from
Eq.\ (20) augmented by a Fermi liquid correction $\propto T^{2}$. It fits
the data \cite{physicab} well, as shown in Fig.\ 5.

\begin{figure}[h]
\centering
\includegraphics[totalheight=.3\textheight, viewport=100 80 950
600,clip]{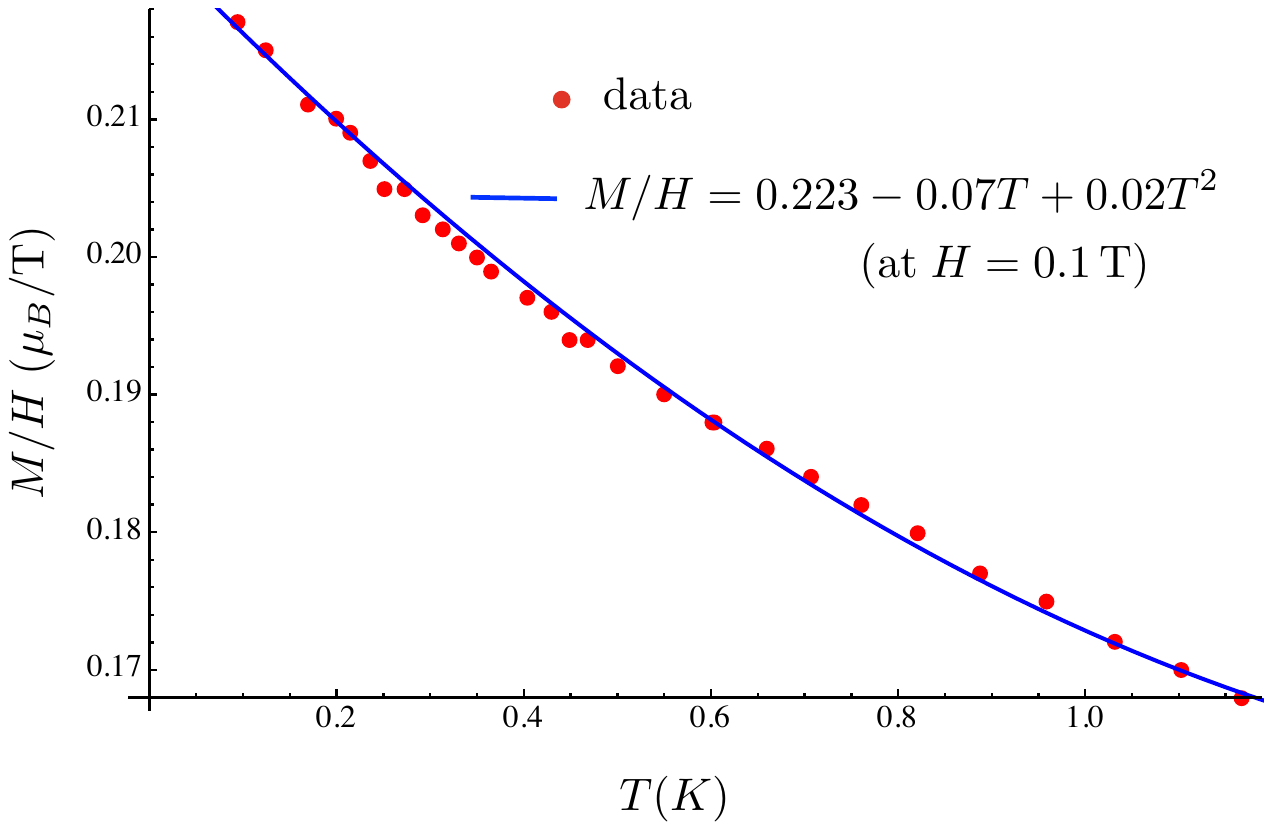} 
\caption{Uniform magnetization: Comparison of theory Eq.\ (20) and
experimental data \protect\cite{physicab} for CeCu$_{6-x}$Au$_x$ at the
critical concentration $x=0.1$}
\label{mag}
\end{figure}



\section{Conclusion}

We presented a semi-phenomenological theory of quantum criticality near a
critical point that separates antiferromagnetic and paramagnetic phases of a
metal. Starting from the assumption that the Landau quasiparticle effective
mass diverges on approaching the critical point, giving rise to critical
quasiparticles, we identified the corresponding renormalization of the
dynamical spin susceptibility near the antiferromagnetic wave vector.
Critical contributions to the electron self energy induced by
antiferromagnetic fluctuations in a clean system are known to be strongly
anisotropic (confined to the \textquotedblleft hot spots"). However,
impurity scattering may be shown \cite{AW,WA} to distribute the effects of
critical scattering all over the Fermi surface. Here we have shown that even
in a clean system, the critical antiferromagnetic fluctuations give rise to
a critical self energy uniformly over the Fermi surface. This is because the
magnetic fluctuations generate energy fluctuations, which diverge in the
long-wavelength limit. The scattering of quasiparticles off these energy
fluctuations leads to a contribution to the effective mass which is
nominally proportional to a positive power of energy, but is strongly
enhanced by factors of the effective mass itself. This leads to a
self-consistency relation for the effective mass (or the quasiparticle $Z$%
-factor), which may have a strong-coupling solution provided the initial
value of $Z$ at the appropriate high-energy scale is sufficiently small (as
may be caused by precursor fluctuations leading to a weakly-diverging
effective mass). While the critical quasiparticles living on
the\textquotedblleft cold" parts of the Fermi surface dominate most of the
observable quantities, the \textquotedblleft hot" quasiparticles may be
shown to be even more singular, \textit{e.g.} in $d=3$, we find the
equivalent of the $Z$-factor exponent $\eta $ to be  $=1/2$. In this context, it is
interesting to observe that the observed critical behavior of CCA depends on
whether the QCP is tuned by varying the Au concentration, the pressure, or
the magnetic field. It is conceivable that in the case of magnetic field
tuning the precursor fluctuations necessary to access the strong coupling
regime are too weak so that the system remains in the weak-coupling regime,
as is apparently observed. A further condition for the applicability of the
self-consistent solution is that the effective dimension of the bosonic
fluctuations, $z+d$, is sufficiently above the upper critical dimension of
the appropriate field theory (\textit{e.g.} $\phi ^{4}$ theory), such that
boson-boson-interaction effects may be neglected.

Application of our theory to the cases of 3D and 2D fluctuations in a
three-dimensional metal leads to a critically diverging effective mass $%
m^{\ast }\propto T^{-\eta }$ with $\eta =1/4$ (3D) and $\eta =1/8$ (2D), in
good agreement with experimental data on the two heavy-fermion compounds YRS
and CCA, where neutron scattering showed the presence of 3D and 2D AFM
fluctuations. Our theory obeys hyperscaling, taking into account the scaling
of the critical fermions. Further comparisons of our theory with
experimental data on YRS and CCA show good agreement. In particular the
universal behavior of the Gr{\"{u}}neisen ratio in the quantum-disordered
regime measured in YRS is in excellent agreement with our result \cite{AW}.

Finally, we emphasize that we assumed the heavy quasiparticles to be robust,
though modified by scattering from critical spin fluctuations. There is no
breakdown of the Kondo effect nor an associated collapse of part of the
Fermi surface in our scenario. Experimental features, such as the crossover
behavior observed in transport properties (and, to a lesser extent in the
thermodynamic quantities) across the ``$T^*$-line" in the $T-H$ phase
diagram of YRS, may be accounted for by a change in quasiparticle scattering
strength associated with thermal activation of the (ESR) spin resonance as
well as by single quasiparticle spin flip scattering \cite{SR2}.

While the good agreement of our theory with experiment across the board is
encouraging, there is a need to put the phenomenological assumptions it
involves on a firm microscopic basis. Work in this direction is in progress.

\appendix*

\section{correlation length}

The control parameter $r$ in Eq.\ (1) that describes the distance to the
critical point has a non-analytic contribution from the irreducible spin
polarization $\Pi (q,\omega )$, which generates the full susceptibility:

\begin{equation}
\chi(q,\omega)= \frac{\Pi(q,\omega)}{1+\Gamma^{\omega}(q)\Pi(q,\omega)},
\end{equation}
where $\Gamma^{\omega}(q)$ is an irreducible quasiparticle-quasihole
scattering vertex that has a minimum at $\bf{q} = \bf{Q}$.
Schematically, 
\begin{equation}
\Pi(Q,\omega_m) = \sum_{k,n} \lambda_Q^2
G(k,\varepsilon_n)G(k+Q,\varepsilon_n+\omega_m),
\end{equation}
where $\lambda_Q$ is the spin vertex part discussed in the text, Eq.\ (1).

The imaginary part of $\Pi$ arises from Landau damping and is renormalized
by the $\lambda_Q$ vertices: 
\begin{equation}
{\rm Im}\,\Pi(Q, \omega) \propto N_0\lambda_Q^2(\omega/v_FQ).
\end{equation}
In contrast, the real part of $\Pi$ is governed by high-energy contributions
and its leading contribution is unrenormalized ${\rm Re}\,\Pi(Q, \omega)
\propto N_0(1+\ldots)$. There is, however, a non-analytic subleading
contribution to ${\rm Re}\,\Pi$, which may be seen from the
Kramers-Kronig transform on ${\rm Im}\,\Pi$ to be of the form 
\[ 
{\rm Re}\,\Pi(Q,0)\propto N_0\lambda_Q^2 T.
\] 
This contribution is responsible for the $T$-dependence of the correlation
length in the critical region: $1/\xi(T) \propto \lambda_Q(T) T^{1/2}$.

We turn now to the dependence of $\xi $ on the tuning field, say $H$. The
scattering vertex $\Gamma ^{\omega }$ is analytic in $H$, but we expect $\Pi
(Q,\omega =0)$ to be non-analytic at the critical field $H_{c}$. This may be
seen by examining the behavior of $\partial \Pi /\partial H$. From Eq.\ (A2),
one sees that this involves a factor $\partial G/\partial H$ that contains a 
$(\bf{q}=0,\omega \rightarrow 0)$ spin vertex, which is $\propto 1/Z$,
from a Ward identity related to particle number and spin conservation.
Outside the critical cone, $Z(H)\propto (H-H_{c})^{\eta z\nu }$, so that
integrating $\partial \Pi /\partial H\propto 1/Z(H)$, we find ${\rm Re}%
\,\Pi (H)\propto (H-H_{c})^{1-\eta z\nu }$. By equating ${\rm Re}\,\Pi $
and $\xi ^{-2}\propto (H-H_{c})^{z\nu }$, we determine the correlation
length exponent as 
\begin{equation}
\nu =\frac{2}{2+z\eta }
\end{equation}

\begin{acknowledgments}
We acknowledge useful discussions with H.\ v.\ L\"{o}hneysen, F. Steglich, J. Thompson, A. Rosch, Q. Si, C.M. Varma, M. Vojta, and especially A.V. Chubukov. Special thanks to Almut Schr\"{o}der, Oliver Stockert and Philipp Gegenwart for sharing some of their experimental data with us. PW thanks the Department of Physics at the University of Wisconsin-Madison for hospitality during a stay as a visiting professor and acknowledges an ICAM senior scientist fellowship. Part of this work was performed during the summers of 2012-13 at the Aspen Center for Physics, which is supported by NSF grant No. PHY-1066293. J.S. and P.W. acknowledge financial support by the Deutsche Forschungsgemeinschaft through grant SCHM 1031/4-1.
\end{acknowledgments}


\begin{thebibliography}{99}
\bibitem{Hertz} J. A. Hertz, Phys. Rev. B \textbf{14}, 1165 (1976).

\bibitem{Millis} A. J. Millis, Phys. Rev. B \textbf{48}, 7183 (1993).

\bibitem{LRVW} H. v. L\"{o}hneysen, A. Rosch, M. Vojta, and P. W\"{o}lfle,
Rev. Mod. Phys. \textbf{79}, 1015 (2007).

\bibitem{HvL1} H. v. L\"{o}hneysen, \textit{et al}, J. Phys.: Condens.
Matter. \textbf{8}, 9689 (1996)


\bibitem{bro} C.\ Stock \textit{et al}, Phys. Rev. Lett. \textbf{109},
127201 (2012).

\bibitem{hart} This is similar to the composite operator discussion in S.A.
Hartnoll, \emph{et al}, Phys. Rev.\ B \textbf{84}, 125115 (2011)

\bibitem{WA} Peter W\"{o}lfle, and Elihu Abrahams, Phys. Rev. B \textbf{84},
041101 (2011).

\bibitem{AW} Elihu Abrahams, and Peter W\"{o}lfle, Proc. Natl. Acad. Sci.
USA \textbf{109}, 3238 (2012).

\bibitem{rosch} A. Rosch, A. Schr\"{o}der, O. Stockert, and H. v.L\"{o}hneysen Phys. Rev. Lett. \textbf{79}, 159 (1997).

\bibitem{abch} Ar. Abanov and A.V. Chubukov, Phys. Rev. Lett. \textbf{84},
5608 (2000); Phys. Rev. Lett. \textbf{93}, 255702 (2004).

\bibitem{Abanov2003}A. Abanov, A. V. Chubukov, and J. Schmalian,
Adv. in Physics \textbf{52}, 119 (2003). 

\bibitem{metsach} Max A. Metlitski and Subir Sachdev, Phys. Rev. B \textbf{82%
}, 075128 (2010).

\bibitem{Si} Q. Si, S. Rabello, K. Ingersent, and J. L. Smith, Nature
(London) \textbf{413}, 804 (2001).

\bibitem{pc} P. Coleman, C. Pepin, Q. Si, and R. Ramazashvili, J. Phys.:
Condens. Matter \textbf{13}, R723 (2001).

\bibitem{SVS} T. Senthil, M. Vojta, and S. Sachdev, Phys. Rev. B \textbf{69}%
, 035111 (2004); M. Vojta, J. Low Temp. Phys. \textbf{161}, 203 (2010).

\bibitem{CeCuAu} Q. Si, J.-X. Zhu, and D. R. Grempel, J. Phys.: Condens.
Matter, \textbf{17}, R1025 (2005).

\bibitem{pep} I. Paul, C. P{\' e}pin and M.R. Norman, Phys. Rev. Lett. 
\textbf{98}, 026402 (2007); Phys. Rev. B \textbf{78}, 035109 (2008); K-S.
Kim and C. P{\' e}pin Phys. Rev. B \textbf{81}, 205108 (2010) and references
therein.

\bibitem{PW} Peter W{\" o}lfle and Elihu Abrahams, Ann. Phys. (Berlin) 
\textbf{523}, 591 (2011).
\bibitem{pw-cmv} P. W{\" o}lfle and C.M. Varma, in preparation
\bibitem{FM} Ref.\ 6 and P.\ Gegenwart\textit{et al}, Phys. Rev. Lett. 
\textbf{89}, 056402 (2002); K.\ Ishida \textit{et al}, Phys. Rev. Lett. 
\textbf{89}, 107202 (2002); C.\ Krellner \textit{et al}, Phys. Rev. Lett. 
\textbf{100}, 066401 (2008).

\bibitem{MC} D. L. Maslov, and A. V. Chubukov, Phys. Rev. B \textbf{81},
045110 (2010).

\bibitem{schro} A. Schr{\"o}der \textit{et al}, Phys.\ Rev.\ Lett.\ \textbf{%
80}, 5623 (1998).

\bibitem{chu} Here, the $T$-dependence of the self energy $\Sigma$ should be
retained, see A. Chubukov \textit{et al}, Phys.\ Rev,\ B \textbf{71}, 205112
(2005).


\bibitem{senthil} The scaling properties of what we call ``critical
quasiparticles" have been analyzed by T. Senthil, Phys. Rev. B \textbf{78},
035103 (2008).

\bibitem{sc} 
Non-Fermi-liquid fixed points with similar power-law behavior were studied early on for Fermi surfaces coupled to gauge bosons by Chakravarty, Norton and Sylju{\aa}sen, Phys.\ Rev.\ Lett.\ {\bf 74}, 1423 (1995).

\bibitem{Pi}It may be shown that the vertex correction factors $Y_{h,g}$ follow from a new type of Ward identity, which leads to $Y^2 \sim 1/Z^2_c$ . (P. W{\" o}lfle, unpublished).

\bibitem{physicab} H. v. L\"{o}hneysen, \textit{et al}, Physica B \textbf{%
223 \& 224}, 471 (1996).

\bibitem{HvL2} H. v. L\"{o}hneysen, private communication.


\bibitem{SR1} Elihu Abrahams and Peter W\"{o}lfle, Phys. Rev. B \textbf{78},
104423 (2008)

\bibitem{SR2} Peter W\"{o}lfle and Elihu Abrahams, Phys. Rev. B \textbf{80},
235112 (2009) and to be published.







\bibitem{CW} Andrey Chubukov and Peter W\"olfle, Phys.\ Rev.\ B {\bf 89}, 045108 (2014).


\bibitem{AL}  L.G. Aslamazov and A.I. Larkin, Soviet Phys.: Solid State {\bf 10}, 875 (1968)
\end{thebibliography}
\end{document}